\documentclass[12pt]{arXiv_class}
\usepackage{mathrsfs}
\usepackage{authblk}

\title{\textbf{Characterization of depolarizing channels using two-photon interference}} 
\author[1,2,*]{Gustavo C. Amaral}
\author[2,**]{G. P. Tempor\~{a}o}
\affil[1]{QuTech and Kavli Institute of Nanoscience, Delft University of Technology\\
          2600 GA Delft, The Netherlands}
\affil[2]{Center for Telecommunications Studies, Pontifical Catholic University of Rio de Janeiro (CETUC/PUC-Rio)\\
          R. Marqu{\^e}s de S{\~a}o Vicente 225, Rio de Janeiro, RJ, Brazil}
\affil[*]{\textit{G.CastrodoAmaral@tudelft.nl, gustavo@opto.cetuc.puc-rio.br}}
\affil[**]{\textit{temporao@puc-rio.br}}

\begin{document}
\maketitle

\begin{abstract}
Depolarization is one of the most important sources of error in a quantum communication link that can be introduced by the quantum channel. Even though standard quantum process tomography can, in theory, be applied to characterize this effect, in most real-world implementations depolarization cannot be distinguished from time-varying unitary transformations, especially when the time scales are much shorter than the detectors response time. In this paper, we introduce a method for distinguishing true depolarization from fast polarization rotations by employing Hong-Ou-Mandel interference. It is shown that the results are independent of the timing resolutions of the photodetectors.
\end{abstract}

\section{Introduction}
\label{intro}
In the vast majority of quantum optical communication schemes, qubits are encoded in pure states of a single degree of freedom of photons, such as the polarization state or time-bin \cite{rubenok2013real,da2013proof}. This allows for coherent superpositions of states, which are in the core of quantum key distribution (QKD) and many other quantum communication applications \cite{gisin2002quantum}. However, the quantum channel - the propagation media that connects the transmitter (Alice) and receiver (Bob) - can introduce decoherence, by coupling the qubits to other degrees of freedom, usually called the "environment"; this process is known as a depolarizing channel \cite{king2003capacity,daffer2004depolarizing}.

Characterization of a depolarizing channel can be performed by employing Quantum Process Tomography (QPT) \cite{mohseni2008quantum}. In its standard form, QPT is comprised of a series of Quantum State Tomography (QST) procedures on $d^2$ linearly independent input states, where $d$ is the dimension of the Hilbert space. In its turn, QST relies on a series of projective measurements performed on the channel output states \cite{leonhardt1995quantum,altepeter2005photonic}. In practical implementations, the measurements are performed by single-photon detectors (SPD), which have a certain uncertainty (timing jitter) in the time of arrival of the photons. This means that, for example, if the quantum channel performs a coupling between the polarization and time-bin degrees of freedom, and both the coherence time of the optical pulses and the introduced time delay between orthogonal states by the depolarizing channel are smaller than the timing jitter of the detectors, then the measurement will automatically "trace over" the time-bin degree of freedom and will not be able to correctly perform the QST.

It is a known fact in quantum physics that there is no way of distinguishing between quantum states represented by the same density operator. Since the reduced density operator resulting from the depolarizing channel can also be described as an incoherent mixture of pure states, this means that QPT is not able to distinguish between a depolarizing channel and a time-varying unitary operation which randomly rotates the input states in a time scale smaller than the SPD jitter. This is a well-characterized effect in optical fibers, where polarization scramblers are employed to mimic the effect of depolarization \cite{yao2002devices}.

In this work, we propose a method for accessing the time degree of freedom that cannot be directly accessed by the detectors. The main idea comprises of performing joint measurements in pairs of photons, each taken at a different time, via Hong-Ou-Mandel (HOM) interference. In other words, by exploiting photon bunching, which is intrinsically dependent on the indistinguishability of the photonic states, the method probes said indistinguishability and overcomes the limitations of QPT. Whenever a depolarizing channel is present, both photons are in the same quantum state and will therefore interfere and produce bunching; on the other hand, if the channel is replaced by a time-varying unitary operation, the two photons will not be in the same polarization state, and bunching will not take place. 

The structure of the paper is as follows. In section 2, the problem of distinguishing between a depolarizing channel and a fast time-varying unitary transformation is introduced. In section 3, a mathematical model of the distinction method is presented, where the single-photon case is considered for simplicity. In Section 4, a practical example of the method employing coherent states is discussed and simulation results are presented. Section 5 concludes the paper.

\section{The Problem of Characterizing Depolarization}
\label{sec:2}
The depolarizing channel is mathematically described by a completely positive, trace-preserving linear map $\mathcal{E}: \mathcal{B}[\mathcal{H}] \rightarrow\mathcal{B}[\mathcal{H}] $, where $\mathcal{H}$ is the Hilbert space corresponding to the qubits and $\mathcal{B}[X]$ is the space of (bounded) linear operators in $X$, given by:
\begin{equation}
\label{eq1}
\mathcal{E}(\rho) = (1-p)\rho + p\frac{1}{2}I
\end{equation}
where $\rho$ is the density operator describing the input state \cite{daffer2004depolarizing}. 

A straightforward interpretation of eq. \ref{eq1} tells us that the input state $\rho$ has a probability $p$ of being replaced by a completely mixed state or a probability $1-p$ of not being affected by the channel. Clearly, the probability $p$ is related to the output degree of polarization (DOP); assuming the input is always a pure polarization state, then $(1-p)$ corresponds to the output's DOP. Note that eq. \ref{eq1}, though correct, does not describe how depolarization takes place. In fact, there are several ways of obtaining a depolarizing channel; indeed, the Principle of Optical Equivalence, an important theorem due to Stokes, states that different incoherent superpositions of wavefields of the same frequency can result in a beam with the same Stokes parameters \cite{brosseau1998fundamentals}. Quantum theory states the same result in a slight different way: the density operator is a complete description of the quantum state, even though it can be written as a (incoherent) sum of other density operators in a non-unique way \cite{aperes}.

Two different approaches will now be presented. The first one introduces "true" depolarization, i.e. incoherence between two orthogonal polarization states, whereas the second does not actually cause depolarization, but is indistinguishable from the first example.

\subsection{Depolarization by time-polarization entanglement}
\label{subsec1}

In this example, the depolarization process is described by a partial trace over the extended Hilbert space comprised of the photon's polarization state and the environment, corresponding in this example to the time-of-arrival degree of freedom:
\begin{equation}
\label{eq2}
\mathcal{E}(\rho) = \textnormal{Tr}_t \left[ U_{\theta,\phi}(\tau) \rho \otimes\ket{0}_t\prescript{}{t}{\bra{0}} U_{\theta,\phi}(\tau)^\dagger \right]
\end{equation}
where $\ket{0}_t \in \mathcal{H}_t$ is the input state's time-bin and $U_{\theta,\phi}(\tau)$ is a unitary operator acting on the joint Hilbert space $\mathcal{H}\otimes \mathcal{H}_t$, defined by:
\begin{equation}
\label{eq3}
\begin{aligned}
U_{\theta,\phi}(\tau)\ket{\theta,\phi}\ket{0}_t = & e^{i\omega\tau}\ket{\theta,\phi}\ket{\tau}_t \\
U_{\theta,\phi}(\tau)\ket{\theta,\phi}^\perp \ket{0}_t = & \ket{\theta,\phi}^\perp\ket{0}_t 
\end{aligned}
\end{equation}
where $\omega$ is the optical frequency, $\ket{\theta,\phi}$ is a generic polarization eigenstate parametrized by angles $\theta$ and $\phi$ in Poincar{\' e} Sphere and $\ket{\psi}^\perp$ is the orthogonal state to $\ket{\psi}$. Eq. \ref{eq3} can be easily interpreted as follows: the unitary transformation $U_{\theta,\phi}(\tau)$ introduces a differential group delay (DGD) of $\tau$ between its two orthogonal eigenstates. This is exactly what happens, for example, in a polarization-maintaining (PM) optical fiber. For $\omega\tau \ll 1$, eq. \ref{eq3} reduces to the effect of a wave plate or polarization controller.

If we assume a pure state $\ket{\phi_{in}} = (\ket{\theta,\phi}+ e^{i\delta}\ket{\theta,\phi}^\perp )/\sqrt{2}$ in the channel input, for an arbitrary phase difference $\delta$, the output DOP, and therefore the probability $p$ in eq. \ref{eq1}, is completely determined by the relationship between the DGD $\tau$ and the photon's coherence time $\tau_c$. In order to see this, we evaluate eq. \ref{eq2} using the unitary operator of eq. \ref{eq3} and the input state $\rho = \ket{\phi_{in}}\bra{\phi_{in}}$. A straightforward calculation yields:
\begin{equation}
\label{eq4}
\mathcal{E}(\rho) = \frac{1}{2}\begin{bmatrix}
           1 & e^{-i\delta '}\prescript{}{t}{\braket{0|\tau}_t} \\
           e^{i\delta '}\prescript{}{t}{\braket{0|\tau}_t}  & 1 \\
         \end{bmatrix}
\end{equation}
where $\delta ' = \delta - \omega\tau$, and the matrix representation is with respect to the $\{\ket{\theta,\phi}, \ket{\theta,\phi}^\perp\}$ basis. From the reduced density operator, we can now calculate the degree of polarization in the output:
\begin{equation}
\label{eq5}
DOP = \sqrt{1-4\text{det}\left[\mathcal{E}(\rho)\right]} = \left|\prescript{}{t}{\braket{0|\tau}_t} \right|
\end{equation}
which clearly shows that the DGD determines the off-diagonal elements of $\mathcal{E}(\rho)$. The exact relationship will depend, of course, on the photon's temporal coherence. For instance, assuming a Gaussian profile on the photon's temporal wavepackets, and defining the coherence time as the standard deviation of the wavepacket, we have:
\begin{equation}
\label{eq6}
\prescript{}{t}{\braket{0|\tau}_t} = \frac{1}{\tau_c\sqrt{\pi}}\int_{-\infty}^{+\infty}e^{-\tfrac{1}{2}[t^2/\tau_c^2]} e^{-\tfrac{1}{2}[(t-\tau)^2/\tau_c^2]}dt =e^{-\tfrac{1}{4}(\tau/\tau_c)^2},
\end{equation}
and, thus:
\begin{equation}
DOP = e^{-\tfrac{1}{4}(\tau/\tau_c)^2}
\end{equation}

\begin{figure}[ht]
\includegraphics[width=0.6\textwidth]{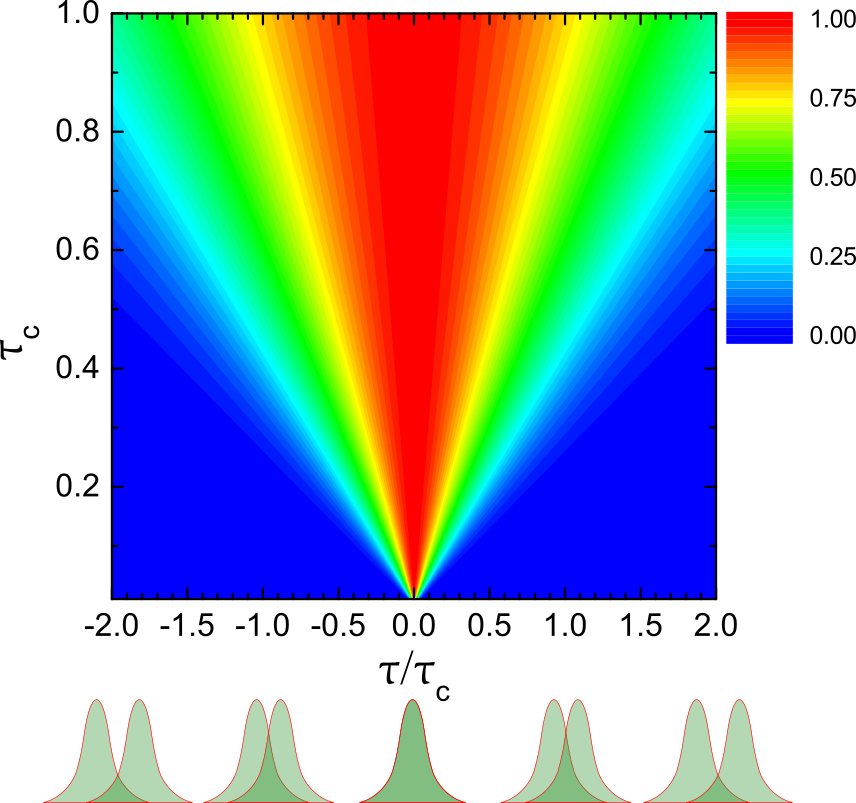}
\caption{Degree of polarization (DOP) as a function of the time-of-arrival mismatch $\tau$ of the wavepackets, assuming gaussian temporal profiles. The red region denotes a high DOP, whereas the blue area indicates low DOP, i.e. high depolarization introduced by the channel.}
\label{fig:1}
\end{figure}

As can be seen from eq. \ref{eq6}, depolarization is achieved whenever $\tau \gg \tau_c$. This can be better visualized in Fig. \ref{fig:1}, for different values of $\tau_c$.

It is important to stress that the reduced density operator of eq. \ref{eq4} actually refers to each one of the photons that go through the quantum channel. All photons are, therefore, identical. Due to the polarization-time coupling, there is no "polarization state" associated to any of the photons. This important remark is what defines this kind of channel as a "true" depolarizing channel.

\subsection{Pseudo-depolarization by incoherent mixing of polarized photons}
\label{subsec2}

Consider now a quantum channel that introduces a random unitary transformation on the input polarization state, i.e. a random rotation in Poincar{\' e} Sphere around the same axis parametrized by angles $(\theta,\phi)$ as in the previous example.  Let $\alpha$ be the rotation angle and $\ket{\phi_{in}} = (\ket{\theta,\phi}+ e^{i\delta}\ket{\theta,\phi}^\perp )/\sqrt{2}$ as before, for any arbitrary phase difference $\delta$. In the $\{\ket{\theta,\phi}, \ket{\theta,\phi}^\perp\}$ basis, the unitary operator that represents the quantum channel is simply given by:
\begin{equation}
\label{eq7}
U(\alpha) = \begin{bmatrix}
           1 & 0\\
           0  & e^{i\alpha} \\
         \end{bmatrix}
\end{equation}
where $\alpha$ is a random variable. Let us assume that it can take only two different values: $-\alpha_0$ and $+\alpha_0$, with equal probability. Then we have, for $\rho = \ket{\phi_{in}}\bra{\phi_{in}}$:
\begin{equation}
\label{eq8}
\begin{aligned}
\mathcal{E}'(\rho) = & \tfrac{1}{2}U(\alpha_0)\ket{\phi_{in}}\bra{\phi_{in}}U^\dagger(\alpha_0) + \tfrac{1}{2}U(-\alpha_0)\ket{\phi_{in}}\bra{\phi_{in}}U^\dagger(-\alpha_0) \\
= & \dfrac{1}{2}\begin{bmatrix}
           1 & e^{-i\delta}\text{cos}(\alpha_0)\\
           e^{i\delta}\text{cos}(\alpha_0)  & 1 \\
         \end{bmatrix}
\end{aligned}
\end{equation}
And, similarly to eq. \ref{eq5}, we obtain the degree of polarization
\begin{equation}
\label{eq9}
DOP' = \sqrt{1-4\text{det}\left[\mathcal{E}'(\rho)\right]} = \left|\text{cos}(\alpha_0)\right|
\end{equation}
which can assume any value between 0 and 1 as expected. The similarities between the density operators given by eqs. \ref{eq4} and \ref{eq8} are self-evident: the role of the DGD in the first example is now played by the rotation angle $\alpha_0$.  However, the individual photons are completely polarized in this case, and they are not all identical to each other, which are fundamental differences from the previous example.

\section{Theoretical Model}
\label{sec:3}

As previously discussed, two quantum states described by the same density operator are indistinguishable from each other, which means that standard Quantum Process Tomography (QPT) techniques are not applicable for distinguishing the quantum state of eq. \ref{eq4} from the one in eq. \ref{eq8}. To overcome this, the measurements must access the total Hilbert space $\mathcal{H}\otimes \mathcal{H}_t$ and not solely the polarization Hilbert space $\mathcal{H}$.

A first and trivial solution is using photodetectors that can resolve the time delay $\tau$ and apply standard QPT. However, depending on the photon coherence time $\tau_c$, this may not be practically feasible. In the case of very short optical pulses, in the time scale of tens of picoseconds or lower, the timing jitter of single-photon detectors (SPD) is not small enough \cite{you2013jitter}, such that the partial trace operation of eq. \ref{eq2} will be inherently performed. 

\begin{figure}[ht]
  \includegraphics[width=\textwidth]{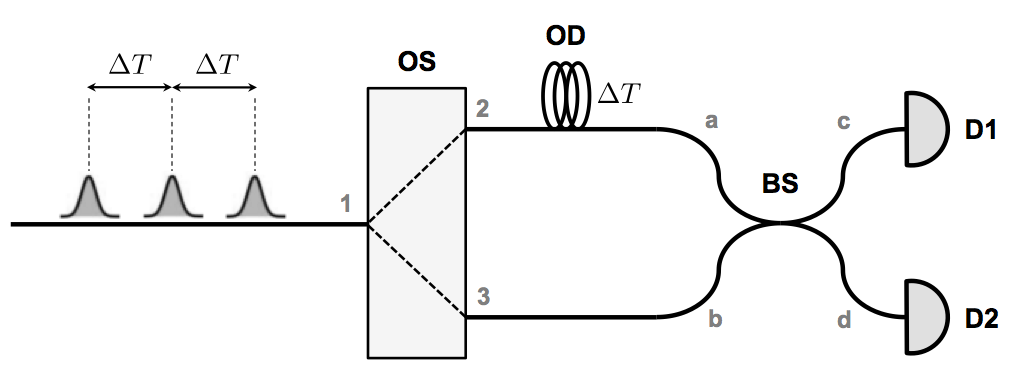}
\caption{Setup for distinguishing true depolarization from an incoherent mixture of pure states employing HOM interference. OS: optical switch; OD: optical delay; BS: beamsplitter; D: single-photon detector}
\label{fig:2}
\end{figure}

We now present a new solution that is completely independent of the SPD timing jitter, which employs Hong-Ou-Mandel (HOM) interference and is shown in Fig. \ref{fig:2}. We assume that the optical fields in the input/output of the quantum channel are comprised of pulsed single photons (i.e. Fock states), with period $\Delta T \gg \tau_c$. An optical switch (OS), with switching frequency of $1/\Delta T$, is coupled to the output of the quantum channel. An optical delay (OD), matched to the pulse period $\Delta T$, is inserted in one of the OS outputs, such that two consecutive photons arrive at the same exact moment in modes $a$ and $b$ of the beamsplitter (BS). 

There are two possible situations for the quantum states at modes $a$ and $b$, corresponding to the quantum channels discussed previously. In the first case (see sects. \ref{subsec1}), both photons in $a$ and $b$ will be in the same quantum state. Let $\ket{0,0}$ represent the two-mode vacuum state and $\hat{k}^\dagger$ be the bosonic creation operator in the spatio-temporal mode defined by $\tau$ and $\tau_c$ in beamsplitter mode $k$. The beamsplitter action can be represented by the pair of unitary operations:
\begin{equation}
\begin{aligned}
\label{eq10}
\hat{a}^\dagger \xrightarrow{\text{BS}} & \tfrac{1}{\sqrt{2}}(\hat{d}^\dagger-i\hat{c}^\dagger) \\
\hat{b}^\dagger \xrightarrow{\text{BS}} & \tfrac{1}{\sqrt{2}}(\hat{c}^\dagger-i\hat{d}^\dagger) \\
\end{aligned}
\end{equation}
If the initial state is given by $\hat{a}^\dagger \hat{b}^\dagger\ket{0,0}$ - one single photon in each input - a simple calculation using eq. \ref{eq10} yields the well-known result, due to HOM interference, of photon bunching:
\begin{equation}
\label{eq11}
\hat{a}^\dagger \hat{b}^\dagger\ket{0,0} \xrightarrow{\text{BS}} \tfrac{1}{\sqrt{2}}\left(\hat{c}^{\dagger 2} + \hat{d}^{\dagger 2}\right)\ket{0,0}
\end{equation}
Therefore, if coincidence measurements are taken between single-photon detectors D1 and D2, no coincidences will be found, i.e. the coincidence rate will be zero. HOM interference does not care whether the quantum states have definite polarization states; the only requirement for eq. \ref{eq10} is that both states are indistinguishable, which also assumes they arrive at the same time in the BS, which is guaranteed by construction.

On the other hand, if the quantum channel is defined by eq. \ref{eq7} (see sect. \ref{subsec2}), there is a probability that photon bunching will not necessarily take place. Without loss of generality, we can assign the horizontal-vertical basis $\{\ket{H},\ket{V}\}$ to the eigenstates of the quantum channel (e.g., by introducing a fixed unitary transformation before the optical switch). In the case the two photons are distinct from each other, we'll have an input state in the BS given by:

\begin{equation}
\label{eq12}
\ket{\text{in}} = \tfrac{1}{\sqrt{2}}\left(\hat{a}^\dagger_{H} + e^{i\alpha_0}\hat{a}^\dagger_{V}\right)\otimes \tfrac{1}{\sqrt{2}}\left(\hat{b}^\dagger_{H} + e^{-i\alpha_0}\hat{b}^\dagger_{V}\right)\ket{0,0}
\end{equation}
where the subscripts in the creation operators indicate the polarization state of each mode. Using eq. \ref{eq10}, a straightforward but somewhat lengthy calculation shows that the state after the BS is given by:
\begin{equation}
\label{eq13}
\begin{aligned}
\ket{\text{out}} = & [\tfrac{1}{2\sqrt{2}}\left(\hat{c}_H^{\dagger 2} +  \hat{c}_V^{\dagger 2} +  \hat{d}_H^{\dagger 2} +  \hat{d}_V^{\dagger 2}\right) - \\ & \tfrac{1}{2}\text{sin}\alpha_0\left(\hat{c}_H^\dagger\hat{d}_V^\dagger - \hat{c}_V^\dagger\hat{d}_H^\dagger\right) - \tfrac{1}{2}\text{cos}\alpha_0\left(\hat{c}_H^\dagger\hat{c}_V^\dagger+ \hat{d}_H^\dagger\hat{d}_V^\dagger\right)] \ket{0,0}
\end{aligned}
\end{equation}
The conditional probability of a coincidence detection given that the two photons are in the joint state described by eq. \ref{eq12} is, therefore, given by:
\begin{equation}
\label{eq14}
P_{coinc} = \tfrac{1}{2}\text{sin}^2\alpha_0
\end{equation}
It is clear that, when $\alpha_0 = 0$, bunching always occurs. However, for $\alpha_0 \neq 0$, there is a nonzero probability of coincidence detections between detectors D1 and D2, which is an indication of the presence of time-varying unitary transformations in the quantum channel. In the case of a completely mixed state ($\alpha_0 = \pi/2$), we have a 50\% probability of coincidence, as expected.

If now we take into account all possible combinations - including the cases where the two photons are in the same quantum state - and also take into account the detection efficiencies of the single-photon detectors and the insertion loss of the optical switch, the probability of generating a coincidence is given by:

\begin{equation}
\label{eq15}
P_{coinc} = \tfrac{1}{4}\eta_{\text{os}}^2\eta_1\eta_2\text{sin}^2\alpha_0
\end{equation}
where $\eta_{\text{os}}$, $\eta_1$ and $\eta_2$ correspond, respectively, to the transmission coefficient of the optical switch and the quantum efficiency of detectors D1 and D2.

Given that a nonzero coincidence probability is only obtained whenever the input states are different from each other, it is possible, using the scheme provided in Fig. \ref{fig:2}, to probabilistically distinguish a depolarizing channel from a random polarization rotation channel with a single measurement. If the measurement results in a coincidence count, then one can conclude, with certainty, that the channel corresponds to the second kind, whereas the absence of a coincidence count does not result in any information gain, i.e., gives an inconclusive result. If the experiment is repeated a large number of times, and all parameters in eq. \ref{eq14} are known, not only one can perfectly distinguish between the two kinds of channels but, additionally, the value of $\alpha_0$ can be estimated.

\section{Practical Scenario -- Discrimination Process with Coherent States}
\label{sec:4}

It should be clear from the reasoning of the previous section that, in case single photons are transmitted through the apparatus of Fig. \ref{fig:2}, a coincidence event will herald the channel's transformation $\mathcal{E}(\rho)$ to be a time-varying unitary operation (Eq. \ref{eq8}). Fock states are, however, impractical for real-life implementations; one could employ Spontaneous Parametric Down-Conversion (SPDC) schemes in order to generate good approximations of pairs of Fock states \cite{gisin2002quantum,Bra_czyk_2010}. A still more practical approach would be employing weak coherent states.

\begin{figure}[ht]
  \includegraphics[width=0.8\textwidth]{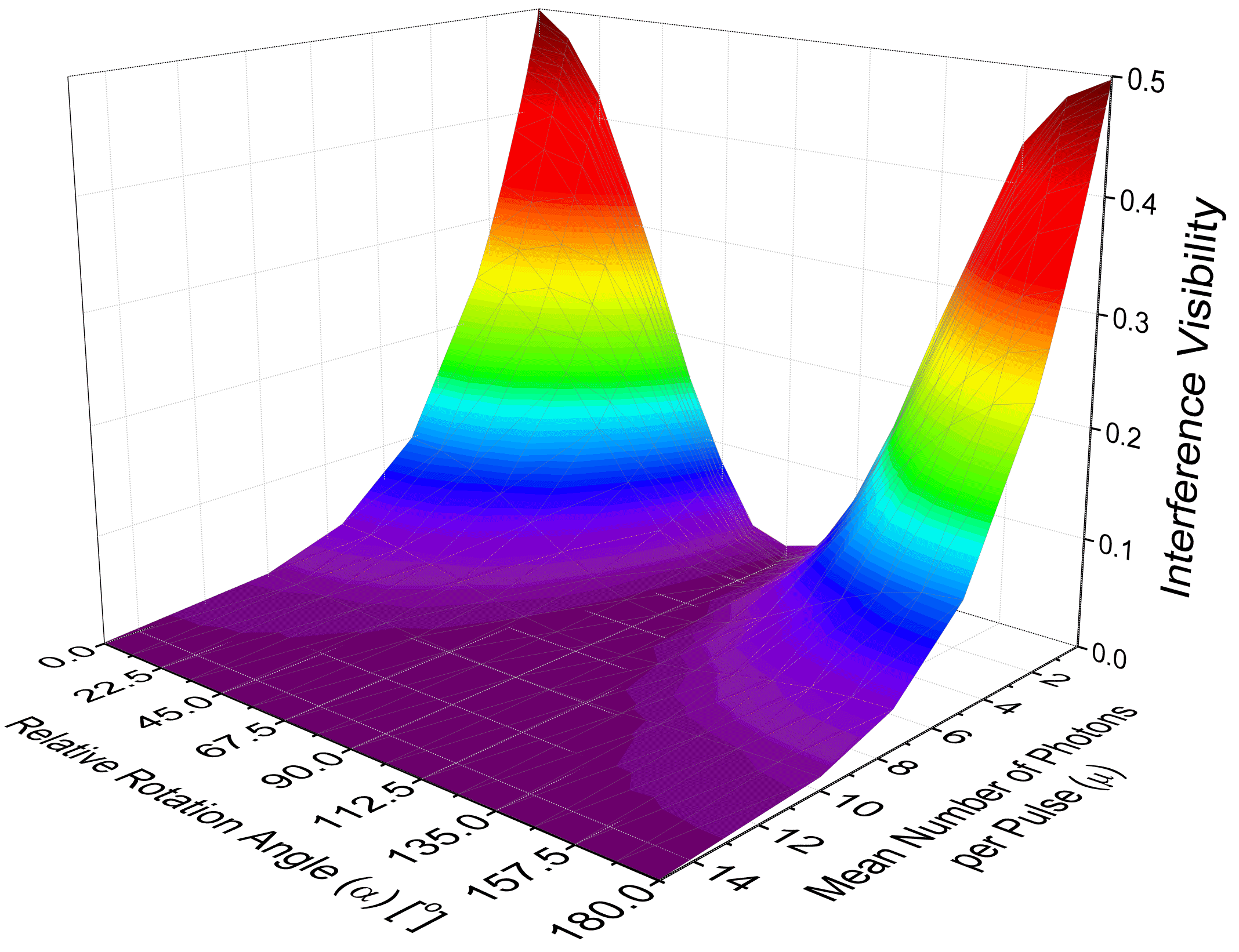}
\caption{Simulation results for Hong-Ou-Mandel visibility as a function of the mean number of photons $\mu$ and the rotation angle $\alpha_0$.}
\label{fig:3}
\end{figure}

Of course, if weak coherent states are employed instead of SPDC-generated photon pairs, the mere presence of a nonzero coincidence rate is no indication whatsoever of time-varying unitary operations. This is because, in the case of coherent states, the multi-photon emission probability introduces an upper bound to the coincidence rate \cite{ou2007multi}. Therefore, instead of simply measuring the coincidence rate between detectors D1 and D2, one now needs to determine the Hong-Ou-Mandel (HOM) visibility of the two-photon interference that takes place in the beamsplitter. It can be shown that, for a mean photon number $\mu \ll 1$, the HOM visibilities for the depolarizing channel of eq. \ref{eq2} and the time-varying unitary rotation of eq. \ref{eq8} are given, respectively, by \cite{moschandreou2018experimental}:
\begin{equation}
\label{eq16}
V_{\text{HOM}}^{\text{depol}} = \tfrac{1}{2}
\end{equation}
\begin{equation}
\label{eq17}
V_{\text{HOM}}^{\text{unit}} = \tfrac{1}{2}\text{cos}^2\alpha_0
\end{equation}
where all imperfections such as dark counts and dead times of the detectors are, initially, not taken into consideration, for simplicity. Fig. \ref{fig:3} depicts the HOM visibility values as a function of $\mu$ and $\alpha_0$. It is possible to observe the effect of the diminishing visibility as the mean photon number per pulse is increased \cite{amaral2018complementarity}.

If we recall the discussion of the single-photon case and apply it to the coherent state case, it is clear that the distinction between the different transformations can only be attested in case the visibility is measured to be different than the maximum expected visibility of $0.5$, which indicates that the channel is introducing a random polarization rotation and $\alpha_0 \neq 0$. In case $\alpha_0 = 0$, the cummulative result will be the maximum expected visibility, which is the same result for the depolarizing channel, i.e., the effect of either cannot be differentiated.

Note that, in order to accurately discriminate the transformation imposed by the channel, a cumulative measurement must be performed such that the uncertainty in the measurement results is smaller than the margin of discrimination; this margin will be given by the difference between the expected visibilities which, in turn, is a function of $\alpha_0$ and the mean number of photons $\mu$ of the incoming weak coherent states. For example, if the HOM visibility is measured as $V=\mathcal{V}+\epsilon$, with $\epsilon$ the uncertainty associated to the measurement, no reliable information can be extracted if $\epsilon > \mathcal{V}-V^{\textrm{max}}$, where $V^{\textrm{max}}$ is the maximum expected visibility. The uncertainty can be decreased at the expense of increased acquisition times; clearly, the lower the mean number of photons, the greater the required acquisition time to obtain a given uncertainty value.

\begin{figure}[ht]
  \includegraphics[width=0.6\textwidth]{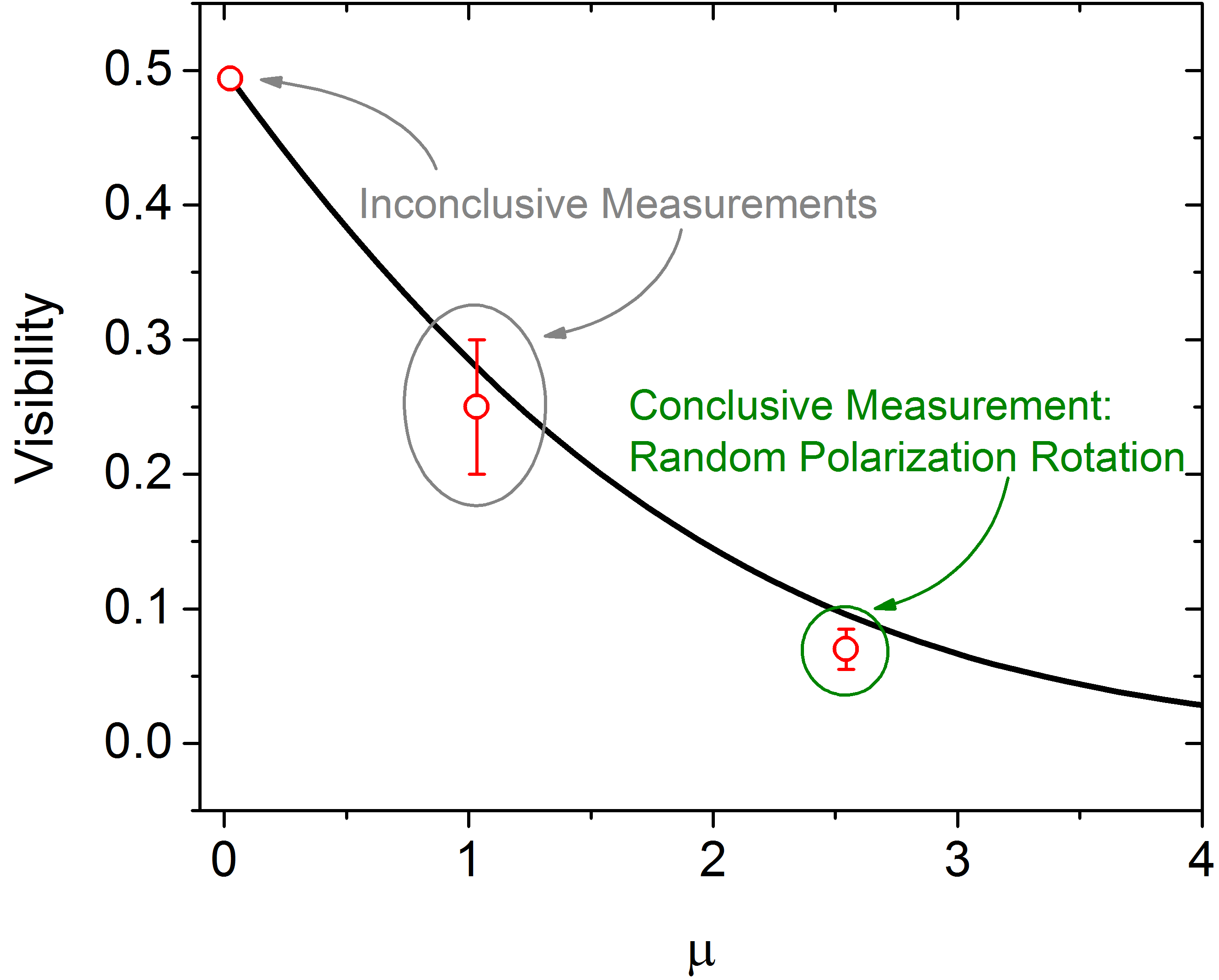}
\caption{Exemplification of three different possible measurements and their associated results. A conclusive result is obtained whenever the measured HOM visibility is lower than the maximum visibility for the corresponding mean number of photons}
\label{fig:4}
\end{figure}

To clarify this discussion, Fig. \ref{fig:4} depicts three distinct measurements that yield conclusive and inconclusive results. The point at $V=0.5$ is inconclusive irrespective of the associated uncertainty, because it can be result of either transformations, since a random polarization rotation with $\alpha_0=0$ (i.e. a fixed unitary transformation) will always achieve maximum visibility. The point at $V\approx0.25$ is also inconclusive, but due to another reason: the uncertainty associated to the measurement is still too high to allow for distinguishing between the two classes of transformations. In this case, more measurements are required in order to reduce the uncertainty. The last point, with $V\approx0.1$ is the only that yields conclusive results, since the uncertainty associated to the measurement is small enough so that one can guarantee that only a random polarization rotation would produce such result.

\section{Conclusion}

The differentiation between transformations imposed by a depolarizing and a pseudo-depolarizing channel has been studied in depth, with a method to practically achieve such distinction being presented for either a single-photon or weak-coherent state case. In the single-photon case, which can be approximated in practice by SPDC sources, a nonzero coincidence rate is sufficient to identify the presence of a time-varying unitary transformation. In case weak coherent states are employed, however, it turns out that a cumulative measurement is necessary in order to make sure that the measurement uncertainty is smaller than the difference between the expected value and the maximum visibility. This fact prompts a practical issue since, on one hand, the greater the number of accumulated coincidence and single counts, the more accurate the results will be (smaller uncertainty); on the other hand, the lower the mean number of photons, the closer to the maximum value the measured visibility will be, but the acquisition time will be exponentially higher.

\section{Acknowledgements}
Financial support from brazilian agencies CNPq, CAPES and FAPERJ is acknowledged.

\end{document}